\documentstyle[twoside,fleqn,espcrc2,epsf]{article}

\newcommand{\AmS}{{\protect\the\textfont2
  A\kern-.1667em\lower.5ex\hbox{M}\kern-.125emS}}

\newcommand{\beqn}{\begin{equation}}
\newcommand{\eqn}{\end{equation}}
\newcommand{\bea}{\begin{eqnarray}}
\newcommand{\eea}{\end{eqnarray}}
\def\lsim{\raise0.3ex\hbox{$<$\kern-0.75em\raise-1.1ex\hbox{$\sim$}}}

\title{
\vspace*{-33pt}
{\normalsize \hfill {\sf UTCCP-P-69}} \\
\vspace*{-6pt}
{\normalsize \hfill {\sf Sept.\ 1999}} \\
Equation of state for SU(3) gauge theory
with RG improved action\thanks{Talk presented by M.~Okamoto
}
}

\author{CP-PACS Collaboration : 
        A.~Ali Khan\rlap, \address{Center for Computational Physics, University of Tsukuba, Tsukuba, Ibaraki 305-8577, Japan}
        S.~Aoki\rlap,\address{Institute of Physics, University of
        Tsukuba, Tsukuba, Ibaraki 305-8571, Japan}
        R.~Burkhalter\rlap,$^{\rm a,b}$
        S.~Ejiri\rlap,$^{\rm a}$
        M.~Fukugita\rlap,\address{Institute for Cosmic Ray Research,
        University of Tokyo, Tanashi, Tokyo 188-8502, Japan}
        S.~Hashimoto\rlap,\address{High Energy Accelerator Research Organization
        (KEK), Tsukuba, Ibaraki 305-0801, Japan}
        N.~Ishizuka\rlap,$^{\rm a,b}$
        Y.~Iwasaki\rlap,$^{\rm a,b}$
        K.~Kanaya\rlap,$^{\rm a,b}$
        T.~Kaneko\rlap,$^{\rm a}$
        Y.~Kuramashi\rlap,$^{\rm d}$
        T.~Manke\rlap,$^{\rm a}$
        K.~Nagai\rlap,$^{\rm a}$
	M.~Okamoto\rlap,$^{\rm b}$
        M.~Okawa\rlap,$^{\rm d}$
        A.~Ukawa\rlap,$^{\rm a,b}$ and
        T.~Yoshi\'e$^{\rm a,b}$ }

\begin{document}

\begin{abstract}
We present results for the equation of state for pure  
SU(3) gauge theory obtained with a renormalization-group (RG)
improved action.
The energy density and pressure are calculated 
on a $16^3\times 4$ and a $32^3\times 8$ lattice employing
the integral method.  Extrapolating the results to the
continuum limit, we find the energy density 
and pressure to be in good agreement with those 
obtained with the standard plaquette action within the 
error of 3--4\%.

\end{abstract}

\maketitle
\setcounter{footnote}{0}
\section{Introduction}

The study of thermodynamic properties of QCD is 
crucial for understanding the early Universe and 
relativistic heavy-ion collisions\cite{reviews}.
The data is encapsulated in the equation of state (EOS). 
Recently the Bielefeld group determined the EOS in the continuum limit 
for pure gauge theory using the standard plaquette action\cite{EOSbielefeld}.
Extending this result to full QCD will require the use of 
improved actions to compensate the increased computer power 
necessary for full QCD simulations. 
As a first step of such a program, we have studied the EOS 
for pure gauge theory 
with a renormalization-group (RG) improved action\cite{Iwasaki83}. 
A summary of results\cite{ourPaper} is presented in this article.

\vspace{-1pt}
\section{Simulation parameters}
\label{sec:simulation}

The RG-improved action we use has the form\cite{Iwasaki83}
$
S_g = c_0\sum( 1\times 1\, {\rm loop}) 
+c_1\sum( 1\times 2\, {\rm loop}) 
$
with $c_0 = 1 - 8c_1$ and $c_1 = -0.331$.

We perform simulations on $16^3\times 4$  and $32^3\times 8$ lattices, 
and also on symmetric $16^4$ and $32^4$ lattices, at about 10 values of 
$\beta=6/g^2$ in the range $T/T_c\approx 0.9$--$3.5$.
We generate $20\,000$ to $36\,000$ iterations after thermalization 
on asymmetric lattices,
and about $10\,000$ iterations on symmetric lattices, 
where one iteration consists of 
one pseudo-heat-bath sweep followed by four over-relaxation sweeps.
Errors are determined by the jack-knife method.

\section{Temperature scale and critical temperature}

We fix the temperature scale using the string tension of the static quark
potential: 
$\frac{T}{T_c} = \frac{(a\sqrt{\sigma})(\beta_c)}{(a\sqrt{\sigma})(\beta)}$.
For this purpose, 
we fit results for $a\sqrt{\sigma}$ \cite{ourPot97} 
using a scaling ansatz proposed by Allton\cite{Allton},
\beqn\label{Allton}
(a\sqrt{\sigma})(\beta) =  
                         f(\beta)  ( \,
  1 + c_2           \, \hat{a}(\beta)^2
    + c_4           \, \hat{a}(\beta)^4 
                            )/ c_0 \,
\eqn
where $f(\beta)$ is the two-loop scaling function of SU(3)
gauge theory, and power corrections in the pseudo lattice spacing 
$\hat{a}(\beta) \equiv { f(\beta) \over f(\beta_1)}$ are introduced
to incorporate deviations from two-loop scaling,
with $\beta_1$ an arbitrary reference point.

\begin{figure}[tb]
\vspace{-5mm}
\begin{center}
\leavevmode
\epsfxsize=7.5cm
\epsfbox{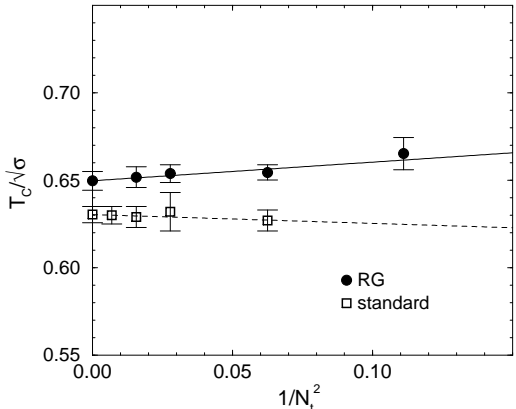}
\end{center}
\vspace{-17mm}
\caption{
$T_c/\protect\sqrt{\sigma}$ as a function of $1/N_t^2$. 
}
\label{fig:Tcscaling}
\vspace{-15pt}
\end{figure}

In Fig.~\ref{fig:Tcscaling}, we plot 
the critical temperature in units of the string
tension, $T_c/\sqrt{\sigma}=1/(N_ta\sqrt{\sigma}(\beta_c))$,
for the RG-improved action
together with the result for the standard plaquette action\cite{Beinlich}.
Making a quadratic extrapolation in $1/N_t$,  
we find $T_c/\sqrt{\sigma}=0.650(5)$ in the 
continuum limit for the RG-improved action, which is 
3\% higher than the value $0.630(5)$ for the standard plaquette 
action\cite{Beinlich}. 
The discrepancy may be caused by
systematic uncertainties in the determination of the string tension
for the two actions, which differ in the details.

\section{Equation of state}
\subsection{Integral method}

We calculate the energy density $\epsilon$ and pressure $p$
using the integral method\cite{Eng90}:
\beqn
\left. {p\over T^4} \right|^{\beta}_{\beta_0} 
= \int^{\beta}_{\beta_0} {\rm d}\beta' \Delta S~,
\;\;\;
{\epsilon - 3p \over T^4} = T{{\rm d}\beta \over {\rm d}T} \Delta S.
\label{eq:intmeas}
\eqn
Here $\Delta 
S \equiv N_t^4 \bigl( \langle S\rangle_T-\langle S\rangle_0\bigr)$,
where $\langle S\rangle_T$ and  $\langle S\rangle_0$ are
the expectation values of the action density $S=S_g/N_s^3 N_t$ 
at finite and zero temperature, respectively.
The beta function $d\beta/dT$ is determined from the scale parametrized 
by (\ref{Allton}).

\begin{figure}[tb]
\vspace{-5mm}
\begin{center}
\leavevmode
\epsfxsize=7.5cm
\epsfbox{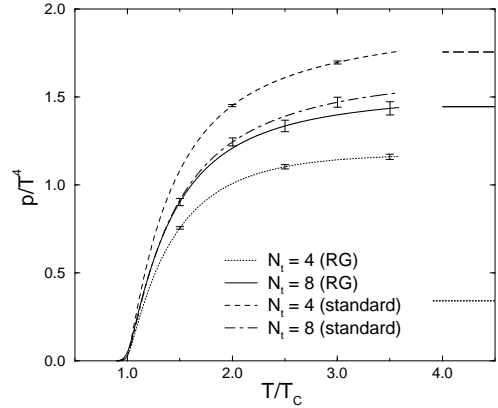}
\end{center}
\vspace{-18mm}
\caption{Pressure for $N_t = 4$ and 8.
}
\label{fig:EOS48}
\vspace{-15pt}
\end{figure}

Our results for the pressure $p$ are shown in Fig.\ref{fig:EOS48}
together with those from the standard action.
We find that the magnitude of the cut-off dependence
for the RG-improved action is similar to that for the standard action,
and opposite in sign.

\begin{figure}[tb]
\vspace{-5mm}
\begin{center}
\leavevmode
\epsfxsize=7.5cm
\epsfbox{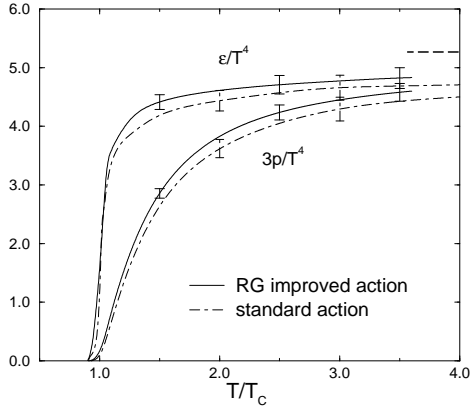}
\end{center}
\vspace{-18mm}
\caption{
Equation of state in the continuum limit for the RG-improved 
action (solid lines) and for the standard  action 
(dash-dotted lines).
}
\label{fig:cont}
\vspace{-15pt}
\end{figure}

We extrapolate the energy density and pressure
to the continuum limit, assuming a quadratic dependence 
on $1/N_t$ as the RG-improved action
has discretization errors of $O(a^2)$.
In Fig.\ref{fig:cont}, results of the extrapolation are plotted 
by solid lines, and are compared with those 
for the standard plaquette action (dot-dashed lines).
As observed in this figure, results in the continuuum limit for the 
two actions are in good agreement with each other within 
the estimated error of 3--4\%.

We note that the curves for the plaquette action in Fig.\ref{fig:cont}
are obtained from a reanalysis of the data in 
Ref.~\cite{EOSbielefeld} employing the same ansatz for the string tension 
(\ref{Allton}) in order to avoid ambiguities arising from the choice 
of scale.  In practice we found the changes in the pressure and 
energy density due to the choice of scale to be very small, being 
less than the statistical error of 1--3\%.

\subsection{Operator method}

The energy density can also be calculated through 
the operator method\cite{Engels82}:
\beqn
\frac{\epsilon}{T^4} =
\frac{18 N_t^4}{g^{2}} \bigl[ c_s (\langle S_s \rangle -
\langle S \rangle_0) -c_t (\langle S_t \rangle
-\langle S \rangle_0) \bigr]
\label{Eope}
\eqn
where $c_s$ and $c_t$ are the asymmetry coefficients.
The pressure can then be obtained with the second equation of 
(\ref{eq:intmeas}).

\begin{figure}[tb]
\vspace{-5mm}
\begin{center}
\leavevmode
\epsfxsize=7.5cm
\epsfbox{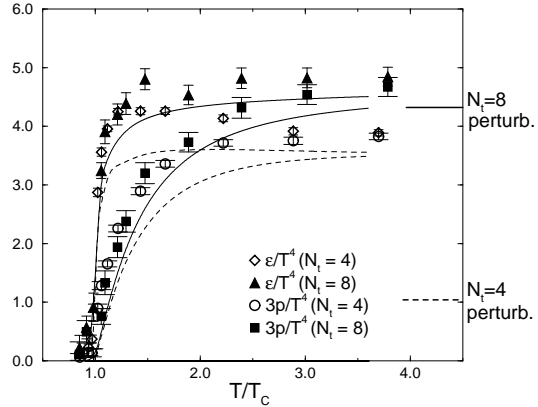}
\end{center}
\vspace{-18mm}
\caption{
EOS from the operator method using one-loop values for 
the asymmetry coefficients
as compared with those from the integral method.
Dashed and solid horizontal lines are
perturbative high temperature limit at $N_t=4$ and 8,
which is common for $\epsilon/T^4$ and $3p/T^4$.
}
\label{fig:operator}
\vspace{-15pt}
\end{figure}

In Fig.~\ref{fig:operator} we compare results obtained 
with the integral and operator methods.
The one-loop perturbative values\cite{Sakai} are used 
for the asymmetry coefficients.
We observe that the two sets of results are consistent with 
each other. The remaining discrepancy may well arise from the use of 
one-loop asymmetry coefficients; 
for the plaquette action, non-perturbative effects are 
known to be important\cite{Ejiri98}.

In the high-temperature limit one may calculate the EOS 
by perturbation theory. The leading-order 
results for the EOS are shown by horizontal lines in Fig.~\ref{fig:operator}.
As has been noted some time ago\cite{Karsch96}, 
the perturbative value for $N_t=4$ is very small for the RG-improved 
action. 
Our numerical results are much larger than this value,
at least in the range $T/T_c$ \lsim\ $3.5$ 
explored in the present work.   

A possible source of the discrepancy is a breakdown of perturbation theory 
due to the infrared divergence.
Another possibility is that pressure and energy density decrease
towards the perturbative values at high temperatures.  This, however, 
would require an unusual situation of a negative $\Delta S$ 
since the pressure is expressed as an integral of $\Delta S$ 
with the integral method.

\section{Conclusions}
Our continuum result for the EOS of pure SU(3) 
gauge theory obtained with an RG-improved gauge action shows 
a good agreement with that from the plaquette action. 
This provides a concrete support for the expectation that continuum 
results are insensitive to the choice of lattice actions.  
We also found that the energy density and pressure for $N_t=4$ 
overshoot the perturbative high temperature limit. 
Understanding the origin of this behavior is left for future investigations. 

\vspace{10pt}

This work is supported in part by the Grants-in-Aid
of Ministry of Education,
Science and Culture (Nos.~09304029, 10640246, 10640248, 10740107, 
11640250, 11640294, 11740162). 
SE, KN and M.\ Okamoto are JSPS Research Fellows.
AAK and TM are supported by the Research for the Future 
Program of JSPS.

\end{document}